\documentclass[epj,referee]{svjour}
\usepackage{epsfig}

\def\bqa{\begin{eqnarray}}
\def\eqa{\end{eqnarray}}
\def\bq{\begin{equation}}
\def\eq{\end{equation}}
\def\dd{{\mathrm d}}
\newcommand{\nll}{\nonumber\\}
\newcommand{\sss}[1]{\scriptscriptstyle{#1}}
\newcommand{\ds }{\displaystyle}

\def\ml {m_{\ell}}
\def\mw {M_{\sss W}}
\def\gw {\Gamma_{\sss W}}
\def\mz {M_{\sss Z}}
\def\gz {\Gamma_{\sss Z}}
\def\mh {M_{\sss H}}
\def\stw{s_{\sss W}}
\def\ctw{c_{\sss W}}
\def\MSbar{\overline{MS}}
\newcommand{\GeV}{\unskip\,\mathrm{GeV}}
\newcommand{\MeV}{\unskip\,\mathrm{MeV}}
\def\hs{\hat s}
\def\hspr{\hat{s}'}
\def\qmo{Q_\ell}
\def\mmo{m_\ell}

\def\qqu{Q_{q}}

\def\chic{\chi^*}
\def\order#1{{\mathcal O}\left(#1\right)}
\def\msbar{\overline{\tiny \mathrm{MS}}}
\def\phAV{\phantom{ AVALUE =}}

\begin{document}

\title{One-loop corrections to the Drell--Yan process in SANC}
\subtitle{(II) The neutral current case.}

\author{A. Arbuzov\inst{1,2} \email{arbuzov@theor.jinr.ru}
\and D. Bardin\inst{2} 
\and S. Bondarenko\inst{1,2} 
\and P. Christova\inst{2} 
\and \\ L. Kalinovskaya\inst{2} 
\and G. Nanava\inst{3}\thanks{on leave from IHEP, TSU, Tbilisi, Georgia}
\and R. Sadykov\inst{2}
}

\institute{Bogoliubov Laboratory of Theoretical Physics, \\
JINR,\ Dubna, \ 141980 \ \  Russia
\and
Dzhelepov Laboratory of Nuclear Problems, \\
JINR,\ Dubna, \ 141980 \ \  Russia
\and
IFJ, \ PAN, \ Krakow, \ 313-42 \ \ Poland}

\abstract{
Radiative corrections to the neutral current Drell--Yan-like processes are considered.
Complete one-loop electroweak corrections are calculated within 
the SANC system. Theoretical uncertainties are discussed.
Numerical results are presented for typical conditions of LHC experiments.
}

\maketitle

\clearpage

\section{Introduction}
 The current theoretical status of Drell--Yan~\cite{Drell:1970wh} physics is widely overviewed 
in the resent papers~\cite{Baur:2007ub,CarloniCalame:2006bg} where the necessity 
of further in-depth study of the Drell--Yan-like (DY) processes is emphasized.

At hadron--hadron colliders DY processes serve for the normalization 
purposes~\cite{Dittmar:1997md,Frixione:2004us}, provide information \linebreak
about weak interactions~\cite{Abazov:2003sv,Abulencia:2005ix},
contribute to the background in many searches for new physics beyond the Standard Model (SM).
At the LHC they will be used also for precision fits of partonic density functions in the regions of $x$ 
and factorization scale values which were not yet accessed experimentally.

 The theoretical calculations of the DY processes for high energy hadronic colliders were performed 
at the level of one-loop QED and electroweak (EW) radiative corrections (RC) by several groups, see papers 
~\cite{Mosolov:1981xk,Soroko:1990ug,Wackeroth:1996hz,Baur:1998kt,Dittmaier:2001ay,Baur:2001ze,Baur:2002fn,Baur:2004ig,CarloniCalame:2006zq,Calame:2007cd} and references therein. 
QCD corrections are known up to the next-to-next-to-leading 
order~\cite{Hamberg:1990np,Anastasiou:2003ds,Melnikov:2006kv}.

 For the experiment, both charge current (CC) and neutral current (NC) DY processes represent great interest. 
Both types of the processes can be easily detected.
CC processes have a larger cross section, but NC processes have a clean dimuon signature and provide an additional 
information, {\it e.g.} about the weak mixing angle.

 The theoretical study for CC DY is pushed ahead more then for NC ---
several independent codes exist for EW NLO RC, see~\cite{CarloniCalame:2006bg}. 
Within the 2005 Les Houches workshop~\cite{Buttar:2006zd} and the 2006 TEV4LHC workshop~\cite{Gerber:2007xk}
a tuned comparison at partonic and hadronic levels at one-loop precision was realized between four codes: 
\cite{Dittmaier:2001ay},~\cite{Baur:2004ig},~\cite{CarloniCalame:2006zq} 
and~\cite{Andonov:2004hi,Bardin:2005dp}.

 To the contrary, a detailed tuned comparison for neutral current DY NLO either at parton or at hadron 
levels is still on the way. In Ref.~\cite{CarloniCalame:2005vc} one can find a comparison between the results 
of the HORACE and ZGRADE codes for the inclusive NC DY cross section at the hadronic level.
Partial results of the comparison at the partonic level between SANC and another 
code can be found in~\cite{Zykunov:2005tc}.

 This article is the second step in the series of SANC papers devoted to DY processes. 
First, we presented in Ref.~\cite{Arbuzov:2005dd} the CC case. Here we show results for the NC case.
Moreover, in the DY branches of SANC we can study the interplay between QCD and EW NLO corrections within 
the same framework. The current status of this study was presented in the report to ATLAS MC working group 
at CERN~\cite{talkKolesnikov:2006}. 
Detailed description of the QCD branch for DY CC and NC processes will be presensed elsewhere, though
the first results are already published in~\cite{Andonov:2006un}.

 The paper is organized as follows. In the second  section we demonstrate the 
implementation of the EW NLO corrections into the SANC framework~\cite{Andonov:2004hi}.
The location of the standard set of the SANC modules: {\bf{FF}} (form factors), 
{\bf{HA}} (helicity amplitudes), 
{\bf BR} and {\bf MC} (bremsstrahlung), is shown on the screenshot of the {\bf SANC Processes}
in the {\bf EW 4f NC} sector.
Ibidem we show the rather short analytical expression for the differential 
cross section of the hard photon contribution.

 SANC team has the advantage of experience in the calculation of bremsstrahlung.
We can obtain the numerical results both by the semi--analytical expressions 
(it is the standard {\bf{BR}} module of the SANC) and by a Monte Carlo integrator 
or generator (with help of the new {\bf{MC}} module).

 In the third  section we describe the adopted ``subtraction'' procedure in the expressions 
for virtual, soft and hard contributions to the NC DY cross section.

 Numerical results are presented in the Sect.~4 both for the Born and for the EW NLO RC.
We investigated the independence of the result on the variation of the soft-hard 
separator $\bar{\omega}$ 
which provides one of the most important checks of the calculation.
The sensitivity of the corrections to the variation of ``subtracted'' quark mass 
singularities was also studied.

 In the conclusion we overview the present status of implementation of NC DY 
calculations into the SANC system.

 As the main point of this article we offer for the import a stand--alone code 
for NC DY EW NLO RC
at the partonic level together with the environment in which it was run.
The sketchy description of this code is presented in Appendix.
Codes are accessible from SANC project homepages 
{\tt http://sanc.jinr.ru} and {\tt http://pcphsanc.cern.ch}.

For production of numbers at the hadronic level the SANC team created a Monte Carlo 
integrator and an event generator, based on the FOAM package~\cite{Jadach:2005ex}. 
The generator itself will be described elsewhere (see~\cite{Sadykov:2007mc}) and made
accessible from the project homepages after intensive rolling and testing.

\section{Neutral current Drell--Yan processes}
\subsection{Born level}
At first we will consider interactions of {\em free} quarks (partons). 
At the leading order (LO) the unpolarized cross section of the partonic subprocess
~$ \bar q(p_1) + q(p_2) \to (\gamma, Z) \to \ell(p_3) + {\bar \ell}(p_4),\,
q=(u,d,c,s,b)$ is given by
\bqa
{\hat\sigma}_0(\hs) =
 \frac{4\, \pi \, \alpha^2}{9 \hs}\, \beta_{\hs} \biggl[
\left(1-\frac{\ml^2}{\hs} \right) V_0(\hs) 
        +\frac{3 \ml^2}{\hs} V_A(\hs)\biggr],\quad
\label{BornNC}
\eqa
where
\bqa
\hs &=&-(p_1+p_2)^2,
\nll
\beta_{\hs}&=&\sqrt{1-\frac{4\ml^2}{\hs}}\,,
\label{sbetal}
\eqa
and $m_\ell$ being the lepton mass. 
Here we denoted
\bqa
 V_{0,A}(\hs) &=& \qqu^2 \qmo^2 + 2\qqu\qmo v_q v_\ell{\rm Re}\chi(\hs) 
\nll 
&&+\left(v_q^2+a_q^2\right)\left(v_\ell^2\pm a_\ell^2\right)\mid \chi(\hs)\mid ^2, 
\nll 
&& v_q = I^{(3)}_q - 2 \qqu \sin^2\theta_W, 
\nll
&& v_\ell = I^{(3)}_\ell - 2 Q_\ell \sin^2\theta_W,
\nll 
&& a_q = I^{(3)}_q,
\nll
&& a_\ell = I^{(3)}_\ell.
\eqa
The $Z/\gamma$ propagator ratio $\chi_{\sss Z}(\hs)$
with constant (or $\hs$-dependent) $Z$ width reads
\bqa
\chi_{\sss{Z}}(\hs)=\frac{1}{4\stw^2\ctw^2}\,\frac{\hs}{\hs-\mz^2+i\mz\Gamma_{\sss Z}}\,,
\eqa
where $\stw$ and $\ctw$ are the sine and cosine of the weak mixing angle $\theta_W$.

\subsection{EW Radiative Corrections at the Partonic Level}
 As the rule of the SANC basement, we subdivide the EW RC into the virtual (loop) ones,
the ones due to soft photon emission, and the ones due to hard photon emission.
Later on in the section~\ref{ResAndComp} in the Tables~\ref{Table1} and~\ref{Table2}, we demonstrate the 
independence of the EW RC off an auxiliary parameter $\bar\omega$ 
which subdivides the soft and hard photonic contributions. 

\begin{figure}[!ht]
\begin{center}
\includegraphics[width=0.4\textwidth]{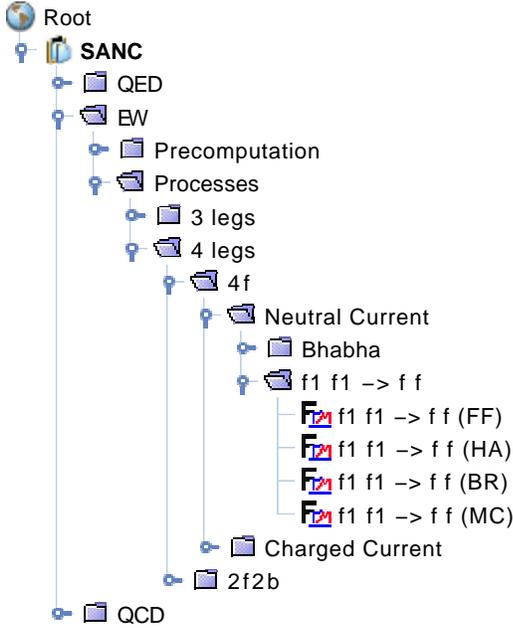}
\end{center}
\caption{SANC tree for the process $2f \to 2f$}
\label{SANCtreeNC}
\end{figure}

 In Fig.~\ref{SANCtreeNC} we show the location of the $2f \to 2f$ NC processes at the SANC tree. 

 Moving along the menu sequence 
{\bf SANC $\to$ EW $\to$ Processes $\to$ 4 legs $\to$ 4f $\to$ 
 Neutral Current $\to$f1~f1~$\to$~f~f~}({\bf FF,HA,BR,MC}) 
user arrives at the first standard SANC module, the scalar Form Factors (FF);
then at the second module, Helicity Amplitudes (HA); then at the third module, 
the integrated bremsstrahlung (BR); and finally at the fourth module, 
the fully differential bremsstrahlung (MC). 

$\bullet$ FF and HA

Earlier, in our article~\cite{Andonov:2004hi} we presented the Covariant (CA) and Helicity Amplitudes (HA) of
 $f_1{\bar f}_1 f{\bar f}\to 0$ NC process, with all 4-momenta being incoming, for any of its cross channels $s,t$ or $u$.
The expressions for the CA and HA
(see Eq.(30) and (33) of the last reference) of this process can be written in terms of six FF.

$\bullet$ BR and MC

The BR module computes the soft and inclusive hard real photon emission: 
\bqa
\bar q(p_1) + q(p_2) \to \ell(p_3) + {\bar \ell}(p_4) + \gamma(p_5), 
\eqa
where the momenta of corresponding particles are given in  brackets.
We do not discuss the soft photon contribution here, referring the reader to 
the system itself. As far as hard photons are concerned, we realized two 
possibilities of the integration over its phase space:
the semi-analytical one (BR) and the one by means of a Monte Carlo integrator 
or generator (MC).

The first one is based on two different sets of analytical integrals over two choices of kinematic 
variables parametrizing the hard photon phase space. 
In the first option (CalcScheme=0), the phase space looks like 
\bqa
d\Phi^{(3)}= d\Phi^{(2)}_1 d\Phi^{(2)}_2 \frac{d\hspr}{2\pi}\,,
\label{PhaseSpase1}
\eqa
where the two-body phase spaces are:
\bqa
d\Phi^{(1)}_1&=&\frac{1}{8\pi}\frac{\sqrt{\lambda(\hs,\hspr,0)}}{\hs}
\frac{1}{2}d\cos{\theta_\gamma}\,,
\nll
d\Phi^{(2)}_2&=&\frac{1}{8\pi}\frac{\sqrt{\lambda(\hspr,m_{\ell}^2,m_{\ell}^2)}}{\hspr}
\frac{1}{4\pi}d\cos{\theta_3}d\phi_3\,.
\label{PhaseSpase11}
\eqa
Here, $\hspr=-(p_3+p_4)^2$ and $\theta_\gamma$ is the cms angle of the photon.
Since, for the chosen parametrization of the phase-space the process matrix element squared does not depend on the 
angle $\phi_{\gamma}$ 
 -- cms azimuthal angle of the photon -- it is already integrated out in the phase-space.
The angles $\theta_3$ and $\phi_3$ 
define the orientation of the momentum $p_3$ in the rest frame of the compound ($\vec{p}_3+\vec{p}_4$=0).
These parameters vary in the limits:
\bqa
&& 0 \leq \theta_\gamma,\theta_3 \leq \pi\,, \quad  0 \leq \phi_3 \leq 2\pi\,,
\nll
&& 4m_{\ell}^2 \leq \hspr \leq \hs-2\sqrt{\hs}\bar{\omega}\,.
\label{sprlimits}
\eqa
After the integration over all angular variables, we obtained the compact expression for a single differential
distribution of the hard photon contribution to the NC DY process,
$\dd\hat{\sigma}_{\mathrm{hard}}/\dd \hspr$.
We neglected terms proportional to the initial quark mass but kept terms proportional to the lepton mass:
\bqa
&&\frac{\dd\hat{\sigma}_{\mathrm{hard}}}{\dd \hspr} = \frac{\alpha}{\pi} \frac{\hs^2+\hspr^2}{\hs^2 \hs_{-}} 
\biggl[ 
  \qqu^2 \left( \ln\frac{\hs}{m^2_q}-1 \right)\hat{\sigma}_0(\hspr)
\nll &&
+ \qmo^2 \left( L'_{\beta} - \beta_{\hspr} \right) 
        \frac{1}{\beta_{\hs}}\hat{\sigma}_0(\hs)
\biggr]
\nll &&
+ \frac{2\alpha^3}{3}\qqu\qmo \frac{\hs_+}{\hs^3}
 \biggl[ A_2(\hs,\hspr) 
 \left(\frac{\mmo^2}{\hspr}\frac{\hs_+}{\hs_{-}}          
                      L'_{\beta}-\frac{\hs}{\hs_-} \beta_{\hspr} \right)
\nll &&
   +\qqu \qmo A_1(\hs,\hspr)\frac{\mmo^2}{\hspr}  L'_{\beta}
 \biggr]
\nll &&
 +\frac{4\alpha^3}{9}\qmo^2\frac{\mmo^2}{\hs^3}
\nll &&
 \times \biggl\{  V_0(\hs)
\biggl[ 2 \left(\frac{\hs_-}{\hs} - 2 \frac{\hs-\mmo^2}{\hs_-}\right) L'_{\beta}
                              - \frac{\hs_-}{\hs} \beta_{\hspr} \biggr]
\nll &&
 \hspace*{1mm} -V_A(\hs) \biggl[ 4 \left(\frac{\hs_-}{\hs} + 3\frac{\mmo^2}{\hs_-}\right) L'_{\beta}
                               - 3 \frac{\hs_-}{\hs} \beta_{\hspr}
         \biggr]
     \biggr\},
\eqa
where
\bqa
L'_{\beta} &=& \ln\frac{1+\beta_{\hspr}}{1-\beta_{\hspr}}\,,
\nll
\hs_{\pm}&=&\hs\pm\hspr\,,
\eqa
with the Born cross section given by Eq.~(\ref{BornNC}). 
The coupling functions are:
\bqa
A_1(\hs,\hspr) &=&
 2 a_q a_\ell{\rm Re}\bigl[\chi(\hspr) - \chi(\hs)\bigr] 
\nll
A_2(\hs,\hspr) &=&
 2\qqu\qmo a_q a_\ell{\rm Re}\bigl[\chi(\hspr)+\chi(\hs)\bigr]
\\
&&+8v_q a_q v_\ell a_\ell{\rm Re}\bigl[\chi(\hs)\chic(\hspr)].
\nonumber
\eqa
We give also a more simple expression neglecting lepton masses. Then all masses remains only 
in the arguments of logarithms:
\bqa
&& \frac{d\hat{\sigma}_{\mathrm{hard}}}{d{\hspr}}= 
   ~\frac{ \alpha}{\pi}\, \frac{1}{\hs_{-}}
  ~\frac{\hs^2+\hspr^2 }{\hs^2}~ 
\nll && 
\times \biggl[
  \qqu^2 \left(\ln\frac{\hs}{m_q^2} - 1 \right)  \hat{\sigma}_0(\hspr)        
+ Q^2_\ell \left(\ln\frac{\hspr}{m_\ell^2} - 1 \right)\hat{\sigma}_0(\hs) \biggr] 
\nll && 
- \frac{1}{3} \,~\frac{{\alpha}^3}{\hs^2}~  
                    \frac{\hs_{+}}{\hs_{-}}
    \,{\mid \qqu Q_\ell \mid} 
    \,{\rm Re}\biggl[  \qqu Q_\ell
         \Bigl( \chi_{\sss Z}(\hs) 
              + \chi_{\sss Z}(\hspr) \Bigr)
\nll &&
  + 4\, v_q v_\ell \, \chi_{\sss Z}(\hs)\chi^{*}_{\sss Z}(\hspr) \biggr],
\eqa
with $\hat{\sigma}_0(\hs)$ being the massless limit of Eq.~(\ref{BornNC}). 

In the second option (CalcScheme=1), the module BR computes the distribution:
$\dd\hat{\sigma}_{\mathrm{hard}}/\dd \hat{c}$, where $\hat{c}=\cos(\angle \vec{p}_2\vec{p}_4)$.
The phase space in this case reads:
\bqa
d\Phi^{(3)}= \frac{1}{2^9\pi^4\,\hs}\,d \hat{c}\,d \hspr\,d Z_4 d\phi_{\gamma}\,,
\label{PhaseSpace2}
\eqa
where $Z_4=-2p_4p_5$ and $\phi_{\gamma}$ is the cms azimuthal angle of $p_5$ varying within $2\pi$ limits.
After integration over $\phi_{\gamma}$, $Z_4$, varying within limits
\bqa
\frac{1}{2}\hs_{-}(1-\beta_{\hs}) \leq & Z_4 & \leq \frac{1}{2}{\hs_{-}}(1+\beta_{\hs}),
\eqa
and $\hspr$ --- within the same limits as in Eq.~(\ref{sprlimits}), one gets very cumbersome result for 
the single differential distribution over variable $\hat{c}$, where one has to neglect terms proportional to all 
masses keeping them only in the arguments of logarithms. The user may see this result after corresponding 
run of the SANC {\bf BR} module with CalcScheme=1.

The {\bf MC} module provides fully differential hard brem\-sstrahlung contribution to the 
partonic cross section. 
The contribution is given in a form suitable for further numerical integration or
simulation of events in a Monte Carlo generator.

\section{Treatment of quark mass singularities \label{subtpro}}
\subsection{Partonic level \label{subtpart}}
 To perform the subtraction procedure at the partonic level cross section
\bq
\hat\sigma_1 = \hat\sigma_0 + \hat\sigma_{\rm SV} + \hat\sigma_{\rm hard}\,,
\label{pSigma1}
\eq
we proceed in the same way as in our previous paper on DY CC~\cite{Arbuzov:2005dd}.

 The subtracted expression $\Delta \hat\sigma^{\msbar}$ from the complete calculations 
with massive quarks:
\bqa \label{sub_part}
\Delta \hat\sigma^{\msbar}_1  
&=& \sum\limits_{i=1,2}^{}
Q_i^2 \frac{\alpha}{2\pi} \int\limits_{\xi_{\rm min}}^1 \dd \xi_i\;  \biggl[
\frac{1+\xi_i^2}{1-\xi_i} \biggl( \ln\frac{M^2}{m_i^2} 
\nonumber \\
&&- 2\ln(1-\xi_i) - 1 \biggr) \biggr]_+ \hat\sigma_0(s\xi_i),
\eqa
where 
\bq
\xi_{\min} = \frac{4\ml^2}{s}\,.
\eq
Next, 
$Q_i$ and $m_i$ are the charge and the mass of the given quark;
$M$ is the factorization scale; $\hat\sigma_0(s\xi_i)$ is the  
cross section at the partonic level with the reduced value 
of $s \to s\xi_i$. 

The plus prescription in Eq.~(\ref{sub_part}) can be treated in the following way:
\bq
 \Delta \hat\sigma^{\msbar}_1 = \lim_{\bar{\omega}\to 0}
\left(\Delta\hat\sigma_{\rm SV} + \Delta\hat\sigma_{\rm hard}\right)^{\msbar}.
\label{pSubst}
\eq

The first contribution, $\Delta\hat\sigma_{\rm SV}$, related to soft and virtual photonic contributions, is given by
\bqa
\Delta\hat\sigma_{\rm SV} &=& \frac{\alpha}{\pi}\qqu^2 
     \biggl( \ln\frac{M^2}{m^2_q}\ln\frac{4\bar{\omega}^2}{s}
          -  \frac{1}{2}\ln^2\frac{4\bar{\omega}^2}{s}
\nll
&&-\ln\frac{4\bar{\omega}^2}{s}+\frac{3}{2}\ln\frac{M^2}{m^2_q} + 2\biggr)\,\hat\sigma_0(s),
\label{DeltaSV}
\eqa
where we took into account that $m_1=m_2=m_q$ and $Q^2_1=Q^2_2=\qqu^2$ for the NC case.

The second one, $\Delta \hat\sigma_{\rm hard}$, related to hard photon emission, is 
\bqa    
\label{DeltaH}
\Delta \hat\sigma_{\rm hard} &=&  \frac{\alpha}{\pi}\qqu^2 
\int\limits_{\xi_{\min}}^{\xi_{\max}} d\xi
\biggl[\frac{1+\xi^2}{1-\xi}\ln\frac{M^2}{\ds m^2_q}
\nll 
&-&  2\ln(1-\xi)-1\biggr]\hat\sigma_{0}(s\xi),
\eqa
where
\bqa 
\xi_{\max} = \frac{s-2\sqrt{s}\bar{\omega}}{s}.
\eqa

Using subtraction procedure, the cross section with $\order{\alpha}$ corrections is given by
\bqa 
\label{msbarsi}
\hat{\sigma}_1^{\msbar} = \hat{\sigma}_1 - \Delta \hat\sigma^{\msbar}_1.  
\eqa
Then it can be convoluted with PDF's in the usual way. An equivalent
subtraction procedure applied at the hadronic level will be discussed right below.

\subsection{Hadronic level \label{subthadr}}
The differential cross section of the DY  process at the hadronic level 
can be obtained from the convolution of the partonic cross section
with the quark density functions:
\bqa 
\label{sigpp}
&& \frac{\dd\sigma_{1}^{pp\to l\bar{l} X}(s,c)}{\dd c} 
= \sum\limits_{q_1q_2}\int\limits_{0}^{1} \int\limits_{0}^{1} 
\dd x_1\; \dd x_2\; \bar{q}_1(x_1,M^2) 
\nonumber \\ && \quad \times
\bar{q}_2(x_2,M^2)
\frac{\dd\hat{\sigma}_1^{q_1\bar{q}_2\to l\bar{l}}(\hat{s},\hat{c})}
{\dd\hat{c}}{\mathcal J}\Theta(c,x_1,x_2),
\eqa
where the step function $\Theta(c,x_1,x_2)$ defines the phase space domain 
corresponding to the given event selection procedure.
The partonic cross section is taken in the center-of-mass reference 
frame of the initial quarks, where the cosine of the muon scattering angle $\hat{c}$ is defined. 
The transformation into the angle $c$ defined in the cms of an initial hadrons involves the Jacobian:
\bqa
&& {\mathcal J} = \frac{\partial \hat{c}}{\partial c}  =
\frac{4x_1x_2}{a^2}\, ,\quad
a = x_1 + x_2 - c(x_1 - x_2), 
\nonumber \\
&& \hat{c} = 1 - (1-c)\frac{2x_1}{a}, \qquad
\hat{s} = sx_1x_2.
\eqa
The parton densities with {\em bars} in Eq.~(\ref{sigpp}) mean the ones
modified by
the subtraction of the quark mass singularities:
\bqa
&& \bar{q}(x,M^2) = q(x,M^2) - \frac{\alpha}{2\pi} \, \qqu^2
\int_x^1 \frac{\dd z}{z} \, q\biggl(\frac{x}{z},M^2\biggr) \, 
\nonumber \\ && \quad \times
\biggl[ \frac{1+z^2}{1-z}
\biggl(\ln\frac{M^2}{m_q^2}-2\ln(1-z)-1\biggr) \biggr]_+\;,
\eqa
where $q(x,M^2)$ can be taken directly from the existing PDF's in the
$\MSbar$
scheme (see Ref.~\cite{Wackeroth:1996hz} for the corresponding formula
in the DIS scheme). It can be shown analytically, 
that this procedure is equivalent to the subtraction from the cross
section, given by Eq.~(\ref{msbarsi}).
In the approach with subtraction from PDF's it is easy to keep the
completely differential form of the sub-process cross section and
therefore to impose any kind of an experimental cut.
When the returning to the $Z$-resonance is allowed by kinematic cuts, 
the {\em natural} choice of the factorization scale is $M^2=\mz^2$. 
For the region of higher invariant masses of the lepton pair,
$M_{l^+l^-}$, it is better to take $M^2\sim M^2_{l^+l^-}$.

\section{Numerical Results\label{ResAndComp}}
The input parameters set is taken the same as used 
in Ref.~\cite{Buttar:2006zd} ({\tt leshw\_input.h}, see Appendix):
\bq 
\begin{array}[b]{lcllcllcl}
G_F & = & 1.16637 \times 10^{-5} \GeV^{-2}, & && \\
\alpha(0) &=& 1/137.03599911, & 
\alpha_s &=& 0.1187, \\
\mw & = & 80.425\GeV, &
\gw & = & 2.124\GeV, \\
\mz & = & 91.1867\GeV,&
\gz & = & 2.4952\GeV,\\
\mh & = & 150\GeV, &
m_t & = & 174.17\;\GeV, \\
m_e & = & 0.51099892\cdot 10^{-3}\;\GeV, &
m_u & = & m_d = 66\;\MeV, \\
m_{\mu}&=&0.105658369\;\GeV, &
m_c & = & 1.55\;\GeV, \\
m_s & = & 150\;\MeV,  & && \\
m_{\tau}&=&1.77699\;\GeV,&
m_b & = & 4.5\;\GeV, \\
|V_{ud}| & = & |V_{cs}| = 0.975, &
|V_{us}| & = & |V_{cd}| = 0.222. 
\end{array}
\label{input}
\eq

\subsection{Partonic level}
We begin with a presentation of several numerical results derived at the partonic level, investigating 
the independence of the results of some unphysical parameters and on the choice of an EW scheme.
First, we investigate the independence off the parameter $\bar{\omega}$, and the residual dependence on 
the initial quark masses after the procedure of subtraction of quark mass singularities 
(see section~\ref{subtpro})
by choosing for quark masses two values: first as in Eq.~\ref{input}, and  the second --- ten times lower.

\subsubsection{Independence off parameter $\bar{\omega}$ and quark masses}
The sum of the ``soft'' and ``hard'' photon contributions to the total and differential cross sections 
should not depend on the soft-hard separator $\bar{\omega}$, 
which we varied from $10^{-3}\frac{\ds\sqrt{\hs}}{\ds 2}$ GeV
to $10^{-5}\frac{\ds\sqrt{\hs}}{\ds 2}$ GeV. We observed that for 
$\bar{\omega}= 10^{-4}\frac{\ds\sqrt{\hs}}{\ds 2}$ 
GeV and $\bar{\omega} = 10^{-5}\frac{\ds\sqrt{\hs}}{\ds 2}$ GeV the numbers for the one-loop corrected cross
sections $\hat{\sigma}_1$ agree within four digits shown, and therefore we present only one row in the 
Tables~\ref{Table1} and~\ref{Table2}. 
For the relative radiative correction factors there is a tiny dependence on $\bar{\omega}$, and we present 
the RC for 
both $\bar{\omega}$'s using marks 1) for $10^{-4}$ and 2) for $10^{-5}$.    
The difference between the results for two $\bar{\omega}$'s is, however, much below any reasonable estimate of the 
theoretical accuracy.
 
The cross sections are shown in picobarn and the radiative corrections in percent.
All the results in Tables~\ref{Table1} and~\ref{Table2} are derived in $\alpha(0)$  EW scheme and
{\em after the subtraction of quark mass singularities in the ${\overline{MS}}$ scheme}.

Each Table contains four rows of $\delta$, two pairs for different $\bar{\omega}$ for $m_u=m_d=0.066$ GeV 
and $m_u=m_d=0.0066$ GeV.
After subtraction of quark mass singularities the total cross section should be fully independent off quark 
masses.
If cuts are applied, a residual dependence arises, in principle, which becomes stronger if cuts get tighter.
The numbers in the Tables~\ref{Table1} and~\ref{Table2} are derived for the only cut $M_{\mu^+\mu^-}\geq$ 50 GeV.
As seen, the residual quark mass dependence is very weak.

All the numbers were obtained with the aid of a FORTRAN code which consists of a hand-written {\tt main } and 
FORTRAN modules automatically generated by {\tt s2n.f} software of the SANC 
system, see Appendix.

\subsubsection{$\alpha(0)$, $G_F$, $G'_{F}$ schemes}
Here we study the EW scheme dependence of the corrected cross section $\hat{\sigma}_1$ 
arising from the definition of EW constants in the $\alpha(0)$, $G_F$ and $G_F^{'}$ schemes.

In Table~\ref{Table3} we show the results for $\hat{\sigma}_0$ and $\hat{\sigma}_1$ at the Born
and one-loop levels and for the corresponding RC factors $\delta$  in three EW schemes.

The $\alpha(0)$ and $G_F$ EW schemes are defined as usually~\cite{Hollik:1985jt}. 
In the $G'_F$ scheme one assigns the same one-loop value of the coupling constant standing at all photon vertices 
$\alpha_{\rm QED}\approx 1/132.544$. It has been adopted in Ref.~\cite{Dittmaier:2001ay} 
and used in Ref.~\cite{Arbuzov:2005dd} for the sake of comparison only.

The results for the radiative corrections are derived in $\MSbar$ subtraction scheme, and 
the factorization scale is taken to be$M_Z$ .
There are notable deviations of the corrected cross sections from the corresponding Born values and between 
the Born values themselves in the three schemes.
One can observe a certain degree of stabilization\hfill of\hfill one-loop \hfill corrected \hfill cross sections as 
 
\clearpage

\begin{table}[!h]
\caption{
\mbox{The total lowest-order and one-loop corrected cross sections 
$\hat{\sigma}_0$ and $\hat{\sigma}_1$ for the process $u\bar{u}\to\mu^+\mu^-(\gamma)$}
\mbox{in the $\alpha$ EW scheme and corresponding relative one-loop correction $\delta$ 
for $\bar{\omega} = 10^{-4}\frac{\ds\sqrt{\hs}}{\ds 2}$ GeV, rows 1) and}
\mbox{ $\bar{\omega} = 10^{-5}\frac{\ds\sqrt{\hs}}{\ds 2}$ GeV, rows 2). Cut value: $M_{\mu^+\mu^-}\geq$ 50 GeV.}
\label{Table1}}
{\bf \small
\begin{tabular}{|c|l|l|l|l|l|l|l|l|}
\hline
$\sqrt{\hs}$, GeV   &          
70 & 90 & 110 & 500 & 1000 & 2000 & 5000 & 14000 \\ \hline
$\hat{\sigma}_0$, pb  
&4.312&369.6&4.995&7.324$\times10^{-2}$&1.806$\times10^{-2}$&4.499$\times10^{-3}$&7.191$\times10^{-4}$&9.171$\times10^{-5}$
\\ \hline   
$\hat{\sigma}_1$, pb     
&4.713&397.2&4.732&8.403$\times10^{-2}$&2.105$\times10^{-2}$&5.256$\times10^{-3}$&8.269$\times10^{-4}$&1.010$\times10^{-4}$   
\\ \hline   
\multicolumn{9}{|c|}{$m_q=m_{u}$}  \\ \hline
$\delta$, $\%$,~1) 
 & 9.303& 7.468&-5.259& 14.74& 16.61& 16.84& 14.99& 10.09
\\ \hline
$\delta$, $\%$,~2)
 & 9.303& 7.481&-5.261& 14.74& 16.61& 16.84& 14.99& 10.09
\\ \hline   
\multicolumn{9}{|c|}{$m_q=m_{u}/10$} \\ \hline
$\delta$, $\%$,~1)
 & 9.296& 7.466&-5.261& 14.74& 16.61& 16.84& 14.99& 10.09
\\ \hline
$\delta$, $\%$,~2)
 & 9.296& 7.480&-5.263& 14.74& 16.61& 16.84& 14.99& 10.09
\\ \hline
\end{tabular}
 }
\end{table}

\begin{table}[!h]
\caption{ 
\mbox{The total lowest-order and one-loop corrected cross sections 
$\hat{\sigma}_0$ and $\hat{\sigma}_1$ for the process $d\bar{d}\to\mu^+\mu^-(\gamma)$}
\mbox{in the $\alpha$ EW scheme and corresponding relative one-loop correction $\delta$ 
for $\bar{\omega} = 10^{-4}\frac{\ds\sqrt{\hs}}{\ds 2}$ GeV, rows 1) and}
\mbox{$\bar{\omega} = 10^{-5}\frac{\ds\sqrt{\hs}}{\ds 2}$ GeV, rows 2). Cut value: $M_{\mu^+\mu^-}\geq$ 50 GeV.}
\label{Table2}}
{\bf \small
\begin{tabular}{|c|l|l|l|l|l|l|l|l|}
\hline
$\sqrt{\hs}$, GeV   &          
70 & 90 & 110 & 500 & 1000 & 2000 & 5000 & 14000 \\ \hline   
$\hat{\sigma}_0$, pb  
 &2.895&472.9&5.237&3.968$\times10^{-2}$&9.601$\times10^{-3}$&2.381$\times10^{-3}$&3.802$\times10^{-4}$&4.847$\times10^{-5}$
\\ \hline   
$\hat{\sigma}_1$, pb    
 &3.079&497.3&5.318&4.439$\times10^{-2}$&1.090$\times10^{-2}$&2.692$\times10^{-3}$&4.149$\times10^{-4}$&4.863$\times10^{-5}$
\\ \hline   
\multicolumn{9}{|c|}{$m_q=m_d$}    \\ \hline
$\delta$, $\%$,~1)
& 6.328& 5.164& 1.551& 11.88& 13.53& 13.04& 9.142& 0.319
\\ \hline
$\delta$, $\%$,~2)
& 6.327& 5.167& 1.550& 11.88& 13.53& 13.04& 9.141& 0.317
\\ \hline
\multicolumn{9}{|c|}{$m_q=m_{d}/10$} \\ \hline
$\delta$, $\%$,~1)
& 6.329& 5.162& 1.552& 11.88& 13.53& 13.04& 9.142& 0.318
\\ \hline
$\delta$, $\%$,~2)
& 6.329& 5.165& 1.551& 11.88& 13.53& 13.04& 9.141& 0.317
\\ \hline
\end{tabular}
 }
\end{table}

\clearpage

\begin{table}[!h]
\caption{
\mbox{The total lowest-order parton cross section $\hat{\sigma}_0$ 
and one-loop corrected cross section $\hat{\sigma}_1$ in pb and relative} 
\mbox{RC factor $\delta$ for the two EW schemes $G_F$ and $G_F'$ for
$\bar{\omega} = 10^{-5}\frac{\ds\sqrt{\hs}}{\ds 2}$ GeV and invariant mass cut $M_{\mu^+\mu^-}$ = 50 GeV;}
\mbox{for details see the text.}
\label{Table3}}
{\bf \small
\begin{tabular}{|c|l|l|l|l|l|l|l|l|}
\hline
$\sqrt{\hs}$, GeV   &          
70     &  90     & 110    & 500    & 1000    & 2000     &  5000    & 14000                 \\ \hline   
\multicolumn{9}{|c|}{$u\bar{u} \to\mu^+\mu^-(\gamma) $}   \\ \hline
\multicolumn{9}{|c|}{$\alpha$}    \\ \hline
$\Delta$ ,$\%$	 
&-0.55 &-1.41    &-0.34   & -0.59  & -0.66   & -0.64    & -0.53    & -0.30
\\ \hline
\multicolumn{9}{|c|}{$G_F$}    \\ \hline   
$\hat{\sigma}_0$, pb  
&4.434&394.9&5.259&7.464$\times10^{-2}$&1.839$\times10^{-2}$&4.581$\times10^{-3}$&7.323$\times10^{-4}$&9.338$\times10^{-5}$ 
\\ \hline
$\hat{\sigma}_1$, pb     
&4.739&402.9&4.748&8.453$\times10^{-2}$&2.119$\times10^{-2}$&5.290$\times10^{-3}$&8.313$\times10^{-4}$&1.013$\times10^{-4}$ \\ \hline
$\delta$, $\%$
&6.878&2.019&-9.718& 13.25            & 15.23               & 15.47              & 13.53              & 8.443
\\ \hline
\multicolumn{9}{|c|}{$G_F'$}    \\ \hline   
$\hat{\sigma}_0$, pb  
&4.608&395.0&5.338&7.826$\times10^{-2}$&1.929$\times10^{-2}$&4.807$\times10^{-3}$&7.685$\times10^{-4}$&9.800$\times10^{-5}$
\\ \hline
$\hat{\sigma}_1$, pb       
&4.805&403.2&4.759&8.657$\times10^{-2}$&2.176$\times10^{-2}$&5.442$\times10^{-3}$&8.576$\times10^{-3}$&1.048$\times10^{-4}$
\\ \hline
$\delta$, $\%$ 
&4.285&2.075&-10.83& 10.61             & 12.76              & 13.21              & 11.60              & 6.913
\\ \hline
$\Delta$, $\%$ 
&+1.01&+0.07& +0.23& +2.41             & +2.69              & +2.87              & +3.16              &+3.46 
\\ \hline
%
%
\hline
\multicolumn{9}{|c|}{$d\bar{d} \to\mu^+\mu^-(\gamma) $}   \\ \hline
\multicolumn{9}{|c|}{$\alpha$}    \\ \hline
&-0.96 &-1.29 &-1.02 & -1.06& -1.09& -1.07&-0.91 &-0.045
\\ \hline
\multicolumn{9}{|c|}{$G_F$}    \\ \hline   
$\hat{\sigma}_0$, pb 
&3.055&505.3&5.573&4.144$\times10^{-2}$&1.002$\times10^{-2}$&2.485$\times10^{-3}$&3.966$\times10^{-4}$&5.057$\times10^{-5}$
\\ \hline
$\hat{\sigma}_1$, pb
&3.109&503.8&5.373&4.487$\times10^{-2}$&1.102$\times10^{-2}$&2.721$\times10^{-3}$&4.187$\times10^{-4}$&4.885$\times10^{-5}$
\\ \hline
$\delta$, $\%$
&1.781&-0.306&-3.593&8.268             & 9.994              & 9.509              & 5.550              &-3.408
\\ \hline
\multicolumn{9}{|c|}{$G_F'$}    \\ \hline   
$\hat{\sigma}_0$, pb 
&3.094&505.3&5.596&4.240$\times10^{-2}$&1.026$\times10^{-2}$&2.545$\times10^{-3}$&4.062$\times10^{-4}$&5.180$\times10^{-5}$
\\ \hline
$\hat{\sigma}_1$, pb
&3.126&503.7&5.371&4.534$\times10^{-2}$&1.115$\times10^{-2}$&2.756$\times10^{-3}$&4.240$\times10^{-4}$&4.940$\times10^{-5}$
\\ \hline
$\delta$, $\%$  
&1.026&-0.310&-4.027& 6.932            & 8.718              & 8.291              & 4.368              & -4.619
\\ \hline
$\Delta$, $\%$  
&+0.55&-0.02 & -0.04&+1.05             &+1.18               &+1.29               & 1.27               & +1.13  
\\ \hline
\end{tabular}
 }
\end{table}

\clearpage

\noindent
can be seen from the rows of $\Delta$ values, showing deviations of $\hat{\sigma}_1$ in 
$\alpha(0)$ and $G_F'$ schemes from $\hat{\sigma}_1$ in $G_F$ scheme:
\bqa
\Delta(\alpha,G'_{F})= \hat{\sigma}_1(\alpha,G'_{F})/\hat{\sigma}_1(G_F)-1,\%\,.
\eqa

However, for the process $u\bar{u}\to\mu^+\mu^-(\gamma)$, especially at high energies, the $G_F'$ 
scheme does not seem to work satisfactorily.

Note, that for $\alpha(0)$ scheme it is sufficient to show only $\Delta$-rows, since all the other entries
are already given in Tables~\ref{Table1} and~\ref{Table2}.

As concerning the choice between the $\alpha(0)$ and $G_F$ schemes, we favor the latter since its effective 
energy scale is known to be of the order of a weak boson mass. 
Moreover, in the results presented in
the $\alpha(0)$ sche\-me, the hadronic contribution to the vacuum polarization 
in the photon propagator is taken into account in an approximation, using
the one-loop formulae supplied by the quark masses listed in Eq.~(\ref{input}).
In any case the difference between the schemes is related to the contribution
of higher orders of the perturbation theory and gives us a crude estimate of 
the corresponding contribution to the theoretical uncertainty.

\subsection{Hadronic level\label{hadlevel}}
Results presented in this section were obtained with the aid of two standalone packages:
1) a MC integrator based on Vegas algorithm~\cite{Lepage:1977sw} and 2) a MC generator based on 
FOAM algorithm~\cite{Jadach:2005ex}.
All the calculations are done with Les Houches 2005 setup~\cite{Buttar:2006zd} in $G_F$ EW scheme.
Besides input parameters Eq.~(\ref{input}), one has to specify cuts on the lepton invariant mass $M_{\mu^{+}\mu^{-}}$,
muons transverse momenta $p_{T}(\mu)$ and muons pseudo-rapidity $\eta({\mu})$ which were used at production of 
figures below:
\bqa
M_{\mu^{+}\mu^{-}}\,&>&\,50\mbox{GeV},
\nll
p_{T}(\mu^{\pm})\,&>&\,25\mbox{GeV},
\nll
|\eta(\mu^{\pm})|\,&<&\,1.2.
\eqa

\begin{figure}[!h]
\includegraphics[width=80mm,height=80mm]{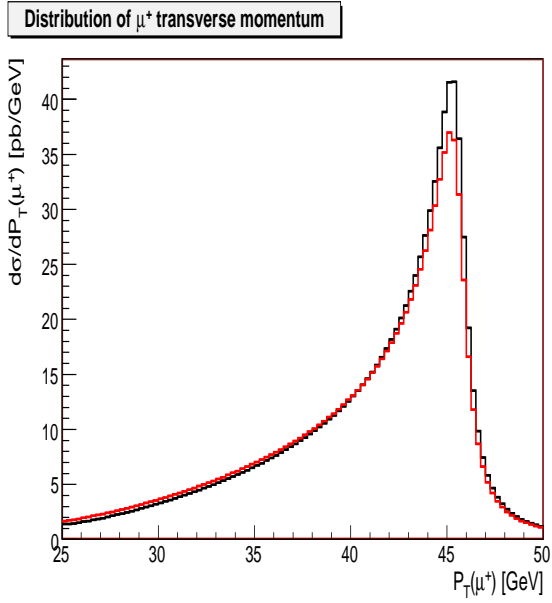}        
\caption{Muon transverse momentum distribution for $\sqrt{s}=14$ TeV, as obtained from the MC integrator.
         Both the Born (black line) and the one-loop results (red line) are shown.
\label{DYsigptint}}
\end{figure}

\begin{figure}[!h]
\includegraphics[width=80mm,height=80mm]{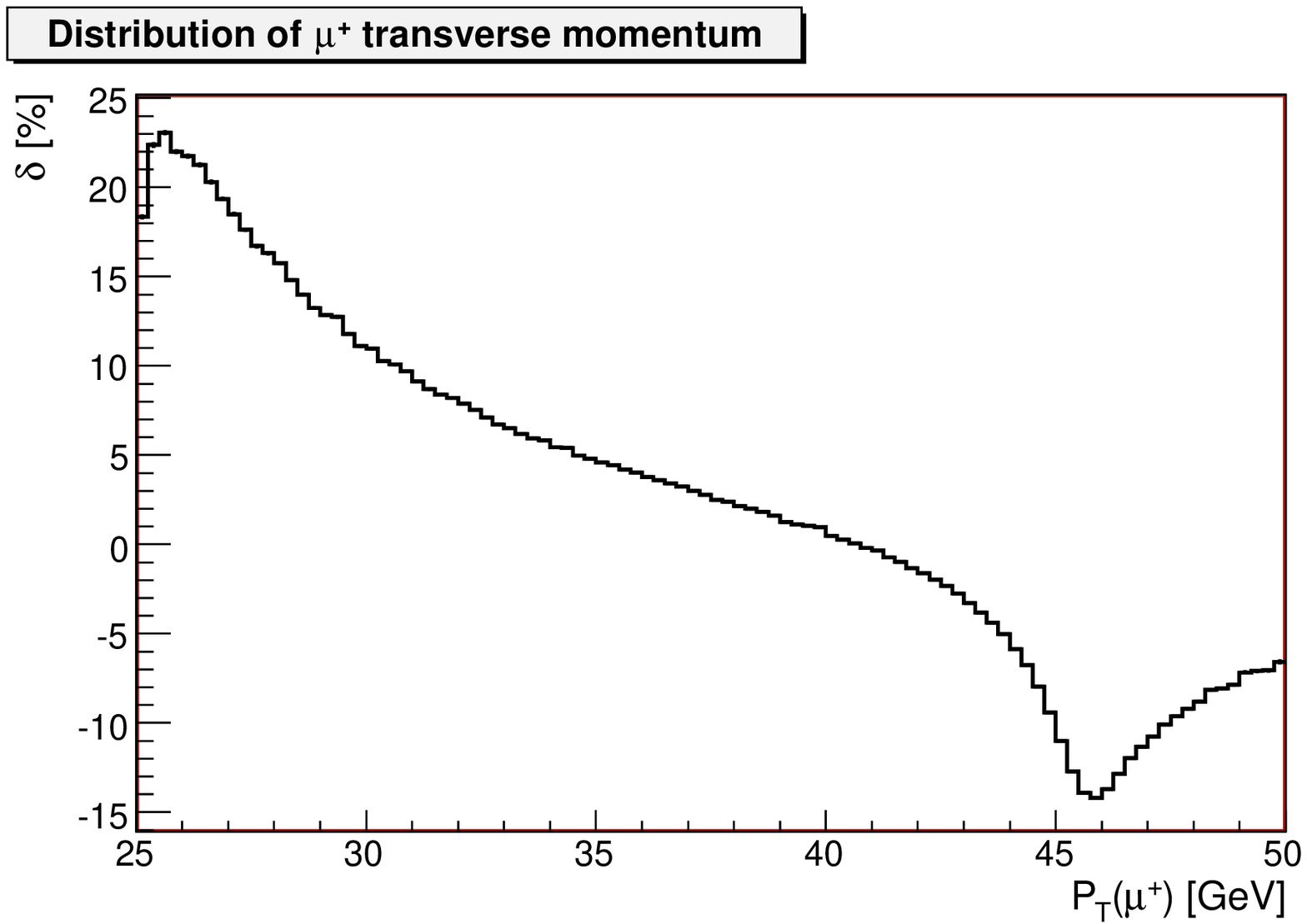}   
\caption{ Relative corrections $\delta$ as a function of the muon transverse momentum $p_{T}(\mu^+)$, 
         as obtained from the MC integrator.
\label{DYdelptint}}
\end{figure}

\begin{figure}[!h]
\includegraphics[width=80mm,height=80mm]{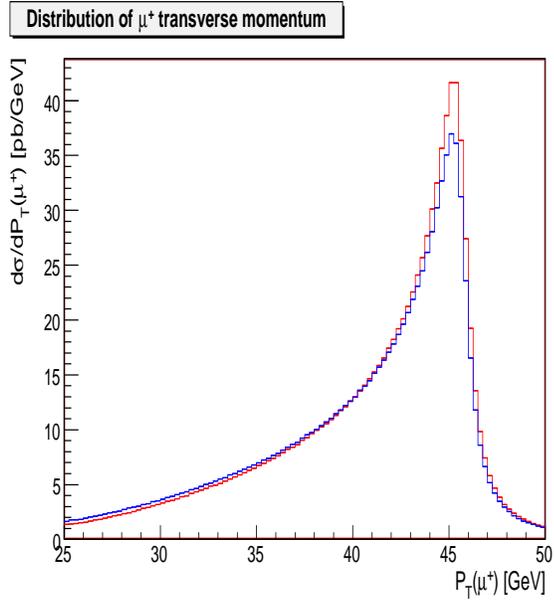}
\caption{Muon transverse momentum distribution for $\sqrt{s}=14$ TeV, as obtained from the MC event generator.
         Both the Born (red line) and the one-loop results (blue line) are shown.
\label{DYsigptgen}}
\end{figure}

\begin{figure}[!h]
\includegraphics[width=80mm,height=80mm]{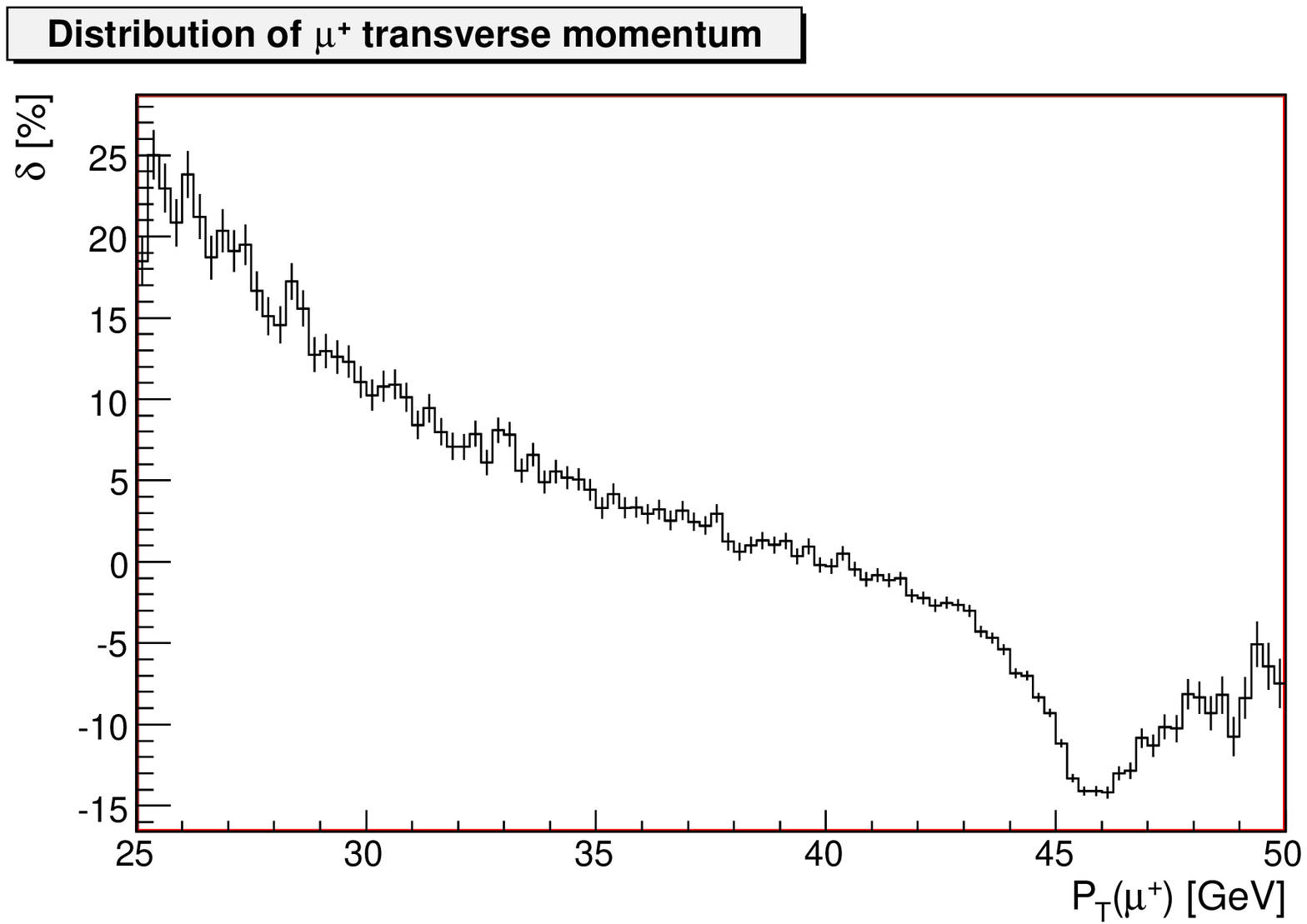}
\caption{Relative corrections $\delta$ as a function of the muon transverse momentum $p_{T}(\mu^+)$, 
         as obtained from the MC event generator.
\label{DYdelptgen}}
\end{figure}

In Figs.~\ref{DYsigptint} and~\ref{DYdelptint} we show the distributions $d{\sigma}/d{p_T}$
and $d\delta/d{p_T}$ obtained with the aid of the MC integrator,
while in the Figs.~\ref{DYsigptgen} and~\ref{DYdelptgen} the same distributions 
obtained with the aid of the generator.
The corresponding distributions agree with each other within statistical errors,
which are larger for the case of the MC event generator.

\begin{figure}[!h]
\includegraphics[width=80mm,height=80mm]{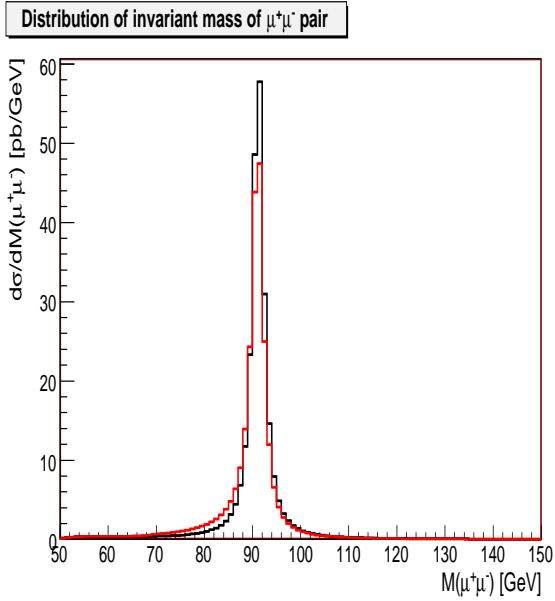}   
\caption{Invariant mass distribution of the $\mu^+\mu^-$ pair at $\sqrt{s}=14$ TeV, as obtained from the MC integrator.
         Both the Born (black line) and the one-loop results (red line) are shown.
 \label{DYsigminvint}}
\end{figure}

\begin{figure}[!h]
\includegraphics[width=80mm,height=80mm]{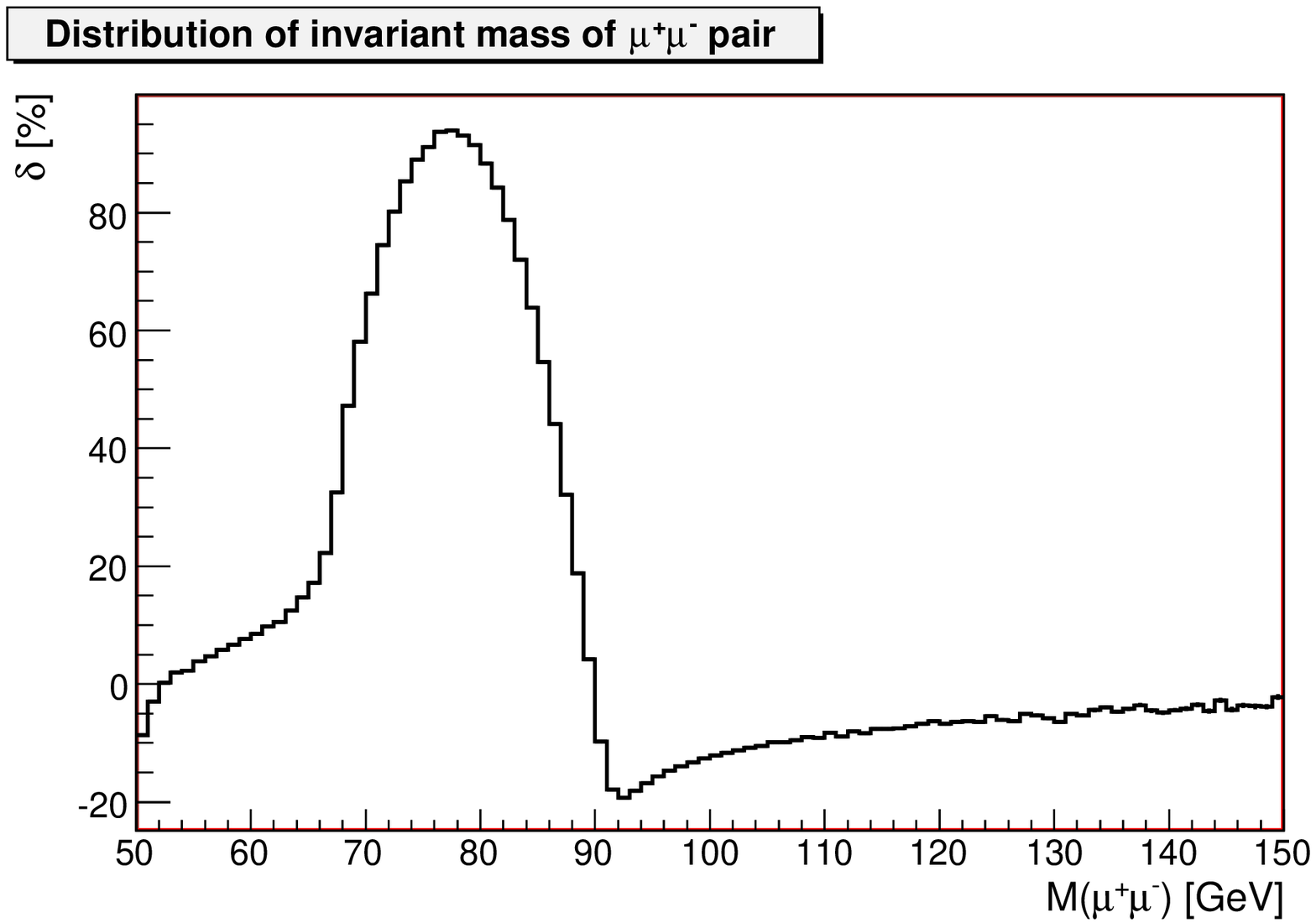}   
\caption{Relative corrections $\delta$ as a function of the invariant mass $M_{\mu^{+}\mu^{-}}$, 
         as obtained from the MC integrator.
\label{DYdelminvint}}
\end{figure}

\begin{figure}[!h]
\includegraphics[width=80mm,height=80mm]{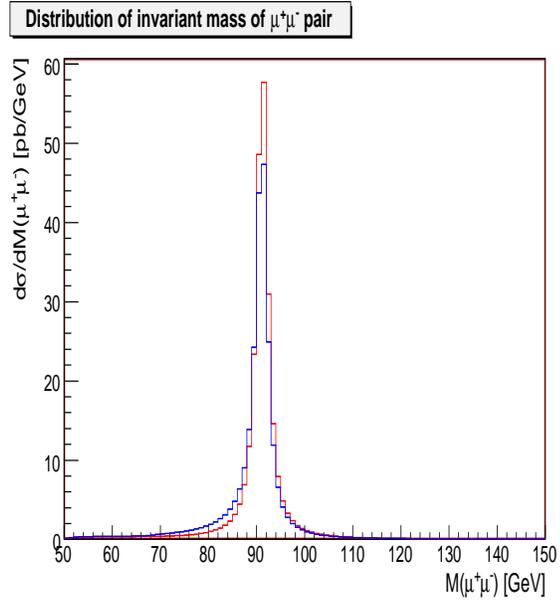}
\caption{Invariant mass distribution of the $\mu^+\mu^-$ pair at $\sqrt{s}=14$ TeV, as obtained from the MC event generator.
         Both the Born (red line) and the one-loop results (blue line) are shown.
  \label{DYsigminvgen}}
\end{figure}

\begin{figure}[!h]
\includegraphics[width=80mm,height=80mm]{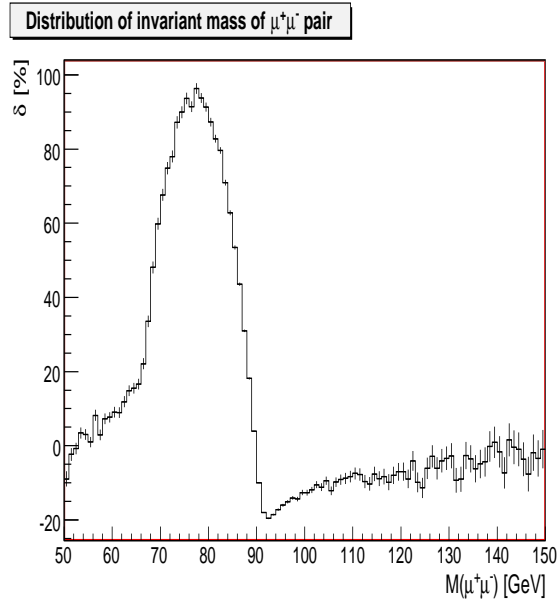}
\caption{Relative corrections $\delta$ as a function of the invariant mass $M_{\mu^{+}\mu^{-}}$, 
         as obtained from the MC event generator.
\label{DYdelminvgen}}
\end{figure}

In the Figures~\ref{DYsigminvint} and~\ref{DYdelminvint} we show the distributions $d{\sigma}/d{M}$ 
and $d\delta/d{M}$ over invariant mass $M=M_{\mu^{+}\mu^{-}}$ for the case of the MC integrator,
and in the Figures~\ref{DYsigminvgen} and~\ref{DYdelminvgen} --- for the case of the ge\-nerator.

One may draw the same conclusions as for the distributions in $p_T$.

The similar distributions are presented in Ref.~\cite{Calame:2007cd} calculated, however, with different
setup and cuts, the\-re\-fore they cannot be compared straightforwardly.

\section{Conclusions}
In addition to the results presented here, one has to take into account
one more $\order{\alpha}$ contribution of the photon--induced
subprocess $\gamma+q\to q + l + \bar{l}$, which was recently computed
in Ref.~\cite{Arbuzov:2007kp}.

In this way the complete set of one-loop electroweak radiative corrections
to neutral current Drell--Yan processes has been computed within the SANC system environment. 
Our results are implemented into semi-analytical {\tt FORTRAN} codes as well as into a Monte Carlo
event generator. 
The codes can help to increase an accuracy of the theoretical description of the SM processes,
which is required for the forthcoming LHC data analysis. 
Tuned comparison of our results for NC DY with the ones of analogous 
calculations performed by other groups, started for the CC case in Refs.~\cite{Buttar:2006zd,Gerber:2007xk}, 
is in progress within the Les Houches 2007 workshop. 
The comparison should help us to derive a common conclusion on the
resulting theoretical uncertainty in our predictions. 

\section*{Appendix\label{appendix}}
In the Appendix we present a Technical Description of the {\tt sanc\_dy\_nc\_v1.00} package intended for
calculation of the total DY NC cross section at the partonic level. 

The main aim of this description is to demonstrate how SANC Standard FORTRAN Modules (SSFM) can be used
by other codes. We also give a short guide to the main flags which are being used not only in this package
but also govern the work of any package which uses SSFM, as our integrators and generators do. 

$\bullet$ {\bf Overview of the package}

 All files are produced by s2n package of SANC project (v.1.10) except
main*\_xx\_yy.F, declaration files --- *.h and libraries --- *.a.
Here ``xx'' and ``yy'' stand for the standard SANC field indices:
$12 - e^{-},\;13 - u,\;14 - d, 16 - \mu^{-}$, {\it etc}.
The package {\tt sanc\_dy\_nc\_v1.00} is accessible for the following set of particles: 
$1313\_1212$, $1313\_1616$, $1414\_1212, 1414\_1616.$

 Total set of files inside the package is:\\
\underline{Instruction files}:
\begin{verbatim}
README
RELEASE-NOTES
CHANGES 
LICENSE.TXT
INSTALL 
\end{verbatim}
\underline{Declaration files}: 
\begin{verbatim}
s2n_declare.h 
\end{verbatim} 
\underline{Initialization and various input files}: 
\begin{verbatim}
s2n_init.f
sanc_input.h
leshw_input.h
tev4lhcw_input.h
\end{verbatim} 
\underline{Libraries for various functions}, including Vegas integration, see INSTALL file in the package.\\
\underline{Main files:} {\tt main\_nc\_vegas\_xx\_yy.F} \\
\underline{SSFM}\hspace{2.5cm} originated from
\begin{verbatim} 					      		   
 nc_ff_xx_yy.F          (FF)
 nc_si_xx_yy.f          (HA)
 nc_br_xx_yy.f          (BR)
[this file contains three SSFM (subroutines)
 nc_bo_xx_yy (...), nc_br_xx_yy (...),             
 nc_ha_xx_xx_1spr (...)]
 nc_ha_xx_yy.f          (MC)
\end{verbatim}
The steps of calculation in the {\tt main*} files in accordance with 
Eqs.~(\ref{msbarsi}), (\ref{pSubst}), (\ref{pSigma1}) are:
\begin{itemize}
\item{}step of \underline{declaration} and \underline{initialization}
\item{}step \underline{born} is realized by flag iborn=1,\\
 ${\hat\sigma_{0}}$ is computed by integration over $\hat{c}$ of Eq.~(\ref{PhaseSpace2})
of the \underline{function$\_$1c$\_$xx$\_$yy},
        \begin{itemize}
        \item{}via SSFM call nc$\_$br$\_$xx$\_$yy (...,born,...)
        \end{itemize}
\item{}step \underline{$\hat\sigma_{\rm SV}$} is realized by flag iborn=0,\\  
it is computed by integration over $\hat{c}$ of \\
the \underline{function$\_$1c$\_$xx$\_$yy},
        \begin{itemize}
        \item{}{\tt virt} via SSFM call nc$\_$si$\_$xx$\_$yy (...,sigma) , \\
              (inside this module there exists a call to SSFM nc$\_$ff$\_$xx$\_$yy (...))
        \item {\tt soft} via SSFM call nc$\_$br$\_$xx$\_$yy (...,soft,...)
        \end{itemize}
\item{}step \underline{$\Delta\hat\sigma_{\rm SV}^{\msbar}$}   \\
   by calculation of $\Delta\hat\sigma_{\rm SV}^{\msbar}$ through Eq.(\ref{DeltaSV})
\item{}step \underline{$\Delta\hat\sigma_{\rm hard}^{\msbar}$} \\
   by integration over $\xi$ of the \underline{function$\_$1s$\_$xx$\_$yy}
         \begin{itemize}
         \item{}via SSFM call nc$\_$bo$\_$xx$\_$yy (...,bornk),\\ see Eq.~(\ref{DeltaH})
         \end{itemize}
\item{step \underline{$\dd\hat{\sigma}_{\mathrm{hard}}/\dd \hspr$}}       \\ 
by integration over $\hspr$ of the \underline{function$\_$1spr$\_$xx$\_$yy}
         \begin{itemize}
         \item{via SSFM call nc$\_$ha$\_$xx$\_$yy$\_$1spr (...,hard)}
         \end{itemize} 
or alternatively
\item{step \underline{$d\sigma_{\rm hard}/d\Phi^{(3)}$}}       \\
 by integration over 4d-phase space of Eq.~(\ref{PhaseSpase1})--(\ref{PhaseSpase11})
of the \underline{function$\_$4d$\_$xx$\_$yy}                       
         \begin{itemize}
         \item{via SSFM call nc$\_$ha$\_$xx$\_$yy (...,hard)}
         \end{itemize}
\end{itemize}

$\bullet$ {\bf Options of the flags}

\begin{description} 
\item[\underline{\bf iqed(I)}] -- choice of calculations for QED correction: 
\item[\phAV I=0]  without~ QED correction
\item[\phAV I=1]  with all QED correction
\item[\phAV I=2]  with ISR QED correction
\item[\phAV I=3]  with IFI QED correction
\item[\phAV I=4]  with FSR QED correction

\end{description}

\begin{description} 
\item[\underline{\bf iew(I)}] -- choice of calculations for EW correction: 
\item[\phAV I=0]  without EW correction
\item[\phAV I=1]  with EW correction
\end{description}

\begin{description} 
\item[\underline{\bf iborn(I)}] -- choice of scheme of calculations of the partonic cross section: 
\item[\phAV I=1]  only Born level
\item[\phAV I=0]  Born + 1-loop virtual corrections
\end{description}

\begin{description} 
\item[\underline{\bf gfscheme(I)}] -- choice of the EW scheme: 
\item[\phAV I=0]  $\alpha$(0)scheme
\item[\phAV I=1]  $G_F $ scheme
\item[\phAV I=2]  $G_F^{'}$ scheme
\end{description}

\begin{description} 
\item[\underline{\bf ilin(I)}] -- choice of the linearization at the calculation of the partonic cross section:
\item[\phAV I=0]  without linearization
\item[\phAV I=1]  {with linearization, {\it i.e.} neglecting \\ \phAV ~ spurious terms ${\cal{O}}(\alpha^2)$}
\end{description}

\begin{description} 
\item[\underline{\bf ifgg(I)}] -- choice of calculations of photonic vacuum polarization ${\cal{F}}_{gg}$:
\item[\phAV I=-1] {--- $0$}
\item[\phAV I= 0] {--- $\rm 1$}
\item[\phAV I= 1] {--- $\rm {1 + k {\cal{F}}_{gg}}$}
\item[\phAV I= 2] {--- $\rm {1/(1 - k {\cal{F}}_{gg})}$}
\end{description}
with $k=\frac{\ds g^2}{\ds 16\pi^2}$.

\begin{description} 
\item[\underline{\bf ihard(I)}] -- types of the hard bremsstrahlung phase-space integrations:
\item[\phAV I=1] {integration over $\hspr$}
\item[\phAV I=4] {4d integration}
\end{description}

\begin{description} 
\item[\underline{\bf isetup(I)}] -- choice of the setup: 
\item[\phAV I=0] {Standard SANC}
\item[\phAV I=1] {Les Houches Workshop}
\item[\phAV I=2] {TeV4LHC Workshop}
\end{description}

The package can be accessed from project homepages
{\it http://pcphsanc.cern.ch/download/sanc\_dy\_nc}\\
{\it \_v1.00.tgz}
{\phantom{a}or\phantom{a}}
{\it http://sanc.jinr.ru/download/sanc\_dy}\\
{\it \_nc\_v1.00.tgz}.

\section*{Acknowledgment}
We are grateful to V.~Kolesnikov, E.~Uglov and V.~Zy\-kunov for discussions.

This work is partly supported by INTAS grant $N^{o}$~03-51-4007,
by the EU grant mTkd-CT-2004-510126 in partnership with the
CERN Physics Department and by the Polish Ministry of Scientific Research and
Information Technology grant No 620/E-77/6.PRUE/ DIE 188/2005-2008 and
by Russian Foundation for Basic Research grant $N^{o}$ 07-02-00932.
One of us (A.A.) thanks also the grant of the President RF Scientific Schools 5332.2006.


\providecommand{\href}[2]{#2}
\end{document}